\documentclass[3p,times,procedia]{elsarticle}
\usepackage{nupha_ecrc}

\volume{00}

\firstpage{1}

\journalname{Nuclear Physics A}

\runauth{E. Andronov}

\jid{nupha}

\jnltitlelogo{Nuclear Physics A}

\usepackage{amssymb}

 \usepackage{lineno}

\usepackage[figuresright]{rotating}

\def\beq{\begin{equation}}
\def\eeq{\end{equation}}

\begin{document}

\begin{frontmatter}



\dochead{XXVIIth International Conference on Ultrarelativistic Nucleus-Nucleus Collisions\\ (Quark Matter 2018)}

\title{Search for the critical point by the NA61/SHINE experiment}


\author{Evgeny Andronov (for the NA61/SHINE Collaboration)}

\address{Saint Petersburg State University, ul. Ulyanovskaya 1, 198504, Petrodvorets, Saint Petersburg, Russia}

\begin{abstract}
NA61/SHINE is a fixed target experiment operating at CERN SPS. Its main goals are to search for the critical point of strongly interacting matter and to study the onset of deconfinement. For these goals a scan of the two dimensional phase diagram ($T$-$\mu_{B}$) is being performed at the SPS by measurements of hadron production in proton-proton, proton-nucleus and nucleus-nucleus interactions as a function of collision energy.

In this paper the status of the search for the critical point of strongly interacting matter by the NA61/SHINE Collaboration is presented including recent results on proton intermittency, strongly intensive fluctuation observables of multiplicity and transverse momentum fluctuations. These measurements are expected to be sensitive to the correlation length and, therefore, have the ability to reveal the existence of the critical point via possible non-monotonic behavior. The new NA61/SHINE results are compared to the model predictions.
\end{abstract}

\begin{keyword}
critical point \sep fluctuations \sep CERN \sep SPS

\end{keyword}

\end{frontmatter}


\section{Introduction}
\label{Introduction}

The NA61/SHINE experiment~\cite{Abgrall:2014fa} is a multi-purpose fixed target experiment at the Super Proton Synchrotron (SPS) of the European Organization for Nuclear Research (CERN). The strong interactions programme of NA61/SHINE is devoted to the studies of the Quantum Chromodynamics phase structure and, in particular, search for the critical point (CP)~\cite{Fodor:2004nz} of strongly interacting matter. 
NA61/SHINE is probing different regions of the phase diagram by performing measurements of hadron production in collisions of protons and various nuclei (p+p, p+Pb, Be+Be, Ar+Sc, Xe+La, Pb+Pb) in a range of beam momenta (13A - 150/158{\it A} GeV/{\it c}). It is expected that there will be a non-monotonic dependence of fluctuations of a number of observables on energy and system size in this scan due to the phase transition of strongly interacting matter and the possible existence of the CP~\cite{Stephanov:1999zu}. Some hints of such behaviour have already been observed by the NA49 experiment~\cite{Grebieszkow:2009jr}.
\section{Multiplicity fluctuations}
Study of event-by-event fluctuations is one of the main tools that is applied to search for the critical point. The strength of fluctuations of a given observable can be quantified by the moments of its distribution. In this paper results on multiplicity fluctuations are presented in terms of the scaled variance, $\omega[N]$, that is constructed from the first and second moments of the multiplicity distribution as $\omega[N]=\frac{\langle N^{2}\rangle -{\langle N\rangle}^{2}}{\langle N\rangle}$, 
where $\langle\cdots\rangle$ stands for the averaging over all events. 
The scaled variance is an intensive quantity which does not depend on the volume of the system for an ideal Boltzmann gas in the grand canonical ensemble (GCE) or on the number of sources within models of independent sources such as the wounded nucleon model (WNM)~\cite{Bialas:1976ed}. 
 
One can expect that $\omega[N]$ is sensitive to the existence of the CP as within the GCE it is proportional to the isothermal compressibility that diverges in the vicinity of the CP~\cite{Mukherjee:2017}, although this signal may be shadowed by significant contribution from volume fluctuations.

The NA61/SHINE experiment obtained preliminary results on multiplicity fluctuations for primary negatively charged hadrons produced in strong and electromagnetic processes in forward energy selected Be+Be (0-5$\%$) and Ar+Sc (0-0.2$\%$) collisions and in inelastic p+p interactions. The selection of the most central events by the forward energy in A+A collisions was done using information from the NA61/SHINE forward calorimeter, the PSD. Transverse momenta of all charged hadrons were restricted to $0<p_{T}<1.5$ GeV/c and rapidities in the center-of-mass system calculated under the pion mass assumption were restricted to $0<y_{\pi}<y_{beam}$. Moreover, the NA61/SHINE acceptance map \cite{ChemFluct_acc} was applied. The results for Be+Be and Ar+Sc collisions were not corrected for detector inefficiencies and trigger biases as simulations have shown that their effect estimated using the GEANT3 package does not exceed 5$\%$. To the contrary the trigger bias modifies results significantly for p+p interactions and, therefore, requires corrections described precisely in Ref.~\cite{Aduszkiewicz:2015jna}. The statistical uncertainties were determined using the sub-sample method. Analysis of the systematic uncertainties is not finished but they are estimated to be smaller than 5$\%$.

\begin{figure}[h]
\begin{center}
\includegraphics[width=12pc]{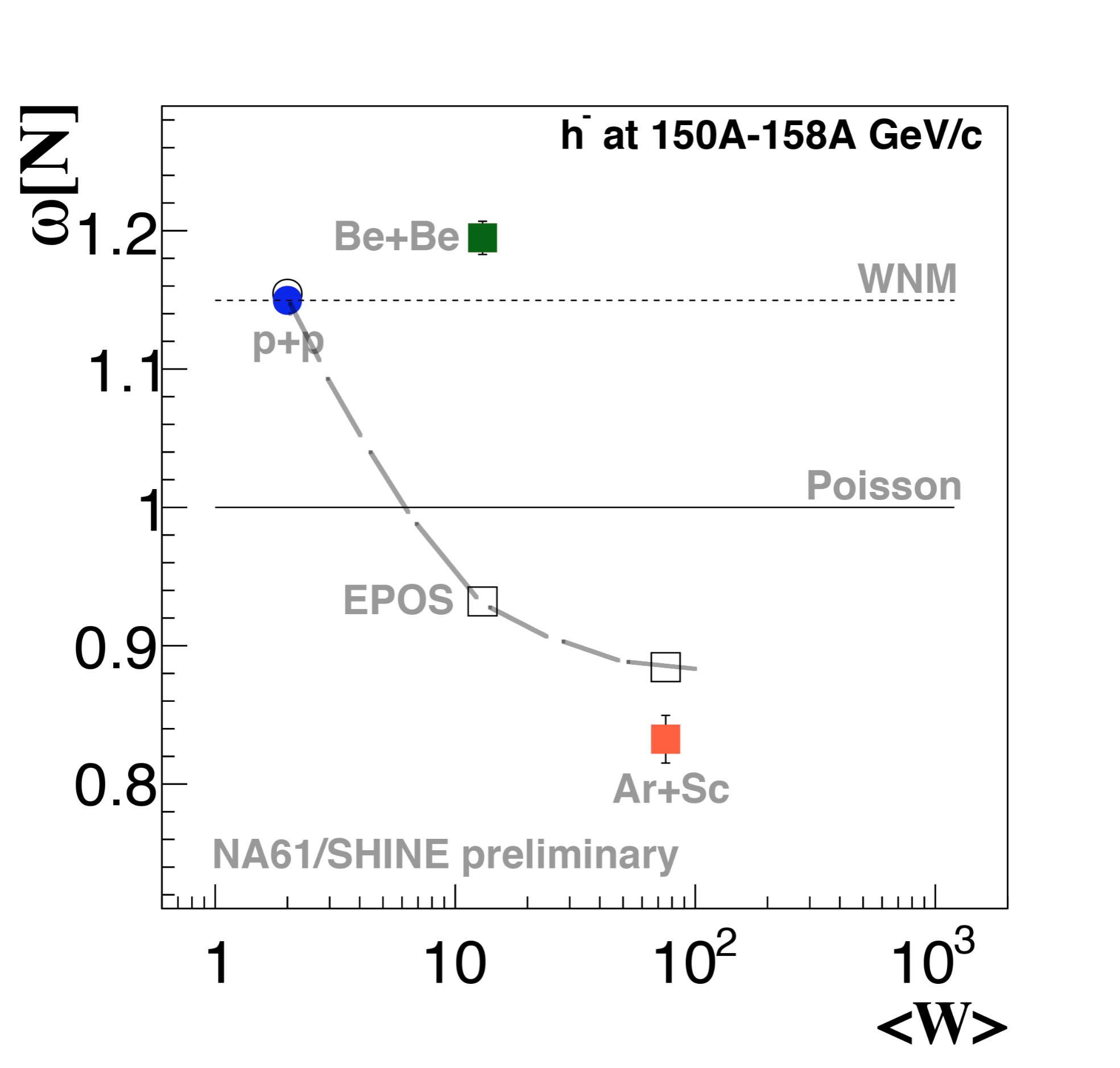}
\end{center}
\caption{\label{omega-fig1} Dependence of $\omega[N]$ for negatively charged hadrons on the mean number of wounded nucleons $\langle W\rangle$ measured by the NA61/SHINE experiment (filled symbols) in comparison to the EPOS1.99 model predictions (open symbols).}
\end{figure}

The results for the dependence of multiplicity fluctuations on the mean number of wounded nucleons, $\langle W\rangle$, are shown in Figure~\ref{omega-fig1}. A dotted line indicates the lower limit for $\omega[N]$ for A+A collisions within the WNM that is defined by the measurement for p+p interactions.  One can see that these model predictions are violated in Ar+Sc collisions. However, $\omega[N]$ obtained for Ar+Sc collisions is close to statistical model predictions for large volume systems~\cite{Begun:2006uu}. Comparison with the EPOS1.99 model predictions~\cite{Pierog:2009zt}, in turn, shows a significant discrepancy in Be+Be collisions. The observed rapid change of $\omega[N]$ when moving from lighter (p+p, Be+Be) to heavier (Ar+Sc) colliding systems can be interpreted as the beginning of creation of large clusters of strongly interacting matter, e.g. as predicted in percolation models~\cite{Celik:1980td,Braun:1991dg}. 

\section{$P_{T}-N$ fluctuation measures}
In order to minimize the influence of volume fluctuations strongly intensive observables are used in search for the critical point \cite{Gorenstein:2011vq, Gazdzicki:2013ana}: 
\begin{eqnarray}
            &\Delta[P_{T},N] = \frac{1}{\langle N\rangle \omega[p_{T}]} \biggl[ \langle N \rangle \omega[P_{T}] -
                        \langle P_{T} \rangle \omega[N] \biggr] \\
            &\Sigma[P_{T},N] = \frac{1}{\langle N\rangle \omega[p_{T}]} \biggl[ \langle N \rangle \omega[P_{T}] +
                        \langle P_{T} \rangle \omega[N] - 2 \bigl( \langle P_{T}N \rangle -
                        \langle P_{T} \rangle \langle N \rangle \bigr) \biggr].
\end{eqnarray}
where $P_{T}=\sum_{i=1}^{N}p_{T_{i}}$ and $\omega[p_{T}]$ is the scaled variance of the inclusive $p_{T}$ spectrum.


In recent measurements by the NA61/SHINE collaboration of the strongly intensive quantities $\Delta[P_{T},N]$ and $\Sigma[P_{T},N]$ in p+p, Be+Be and Ar+Sc collisions no anomaly attributable to the CP was observed~\cite{Andronov:2016ddd}. The analysis was extended to studies of the pseudorapidity dependence of strongly intensive observables in order to probe different values of baryochemical potential~\cite{Brewer:2018abr}. 

In this paper new results were obtained for Be+Be collisions at 150{\it A} GeV/c selected for the smallest 8$\%$ of forward energies. Fluctuations were studied in 9 pseudorapidity intervals defined in the laboratory reference frame. The forward edge of the intervals was fixed at 5.2 units of pseudorapidity, with the backward edge changing from 3 up to 4.6 units. The choice of lower bounds was motivated by the small azimuthal angle acceptance at smaller pseudorapidities. The upper bound was introduced in order to suppress possible quasi-diffractive or electromagnetic effects which become important at larger pseudorapidities. All the other analysis details are identical to the methods described in the previous section. 

\begin{figure}[h]
\begin{center}
\begin{minipage}{12pc}
\includegraphics[width=12pc]{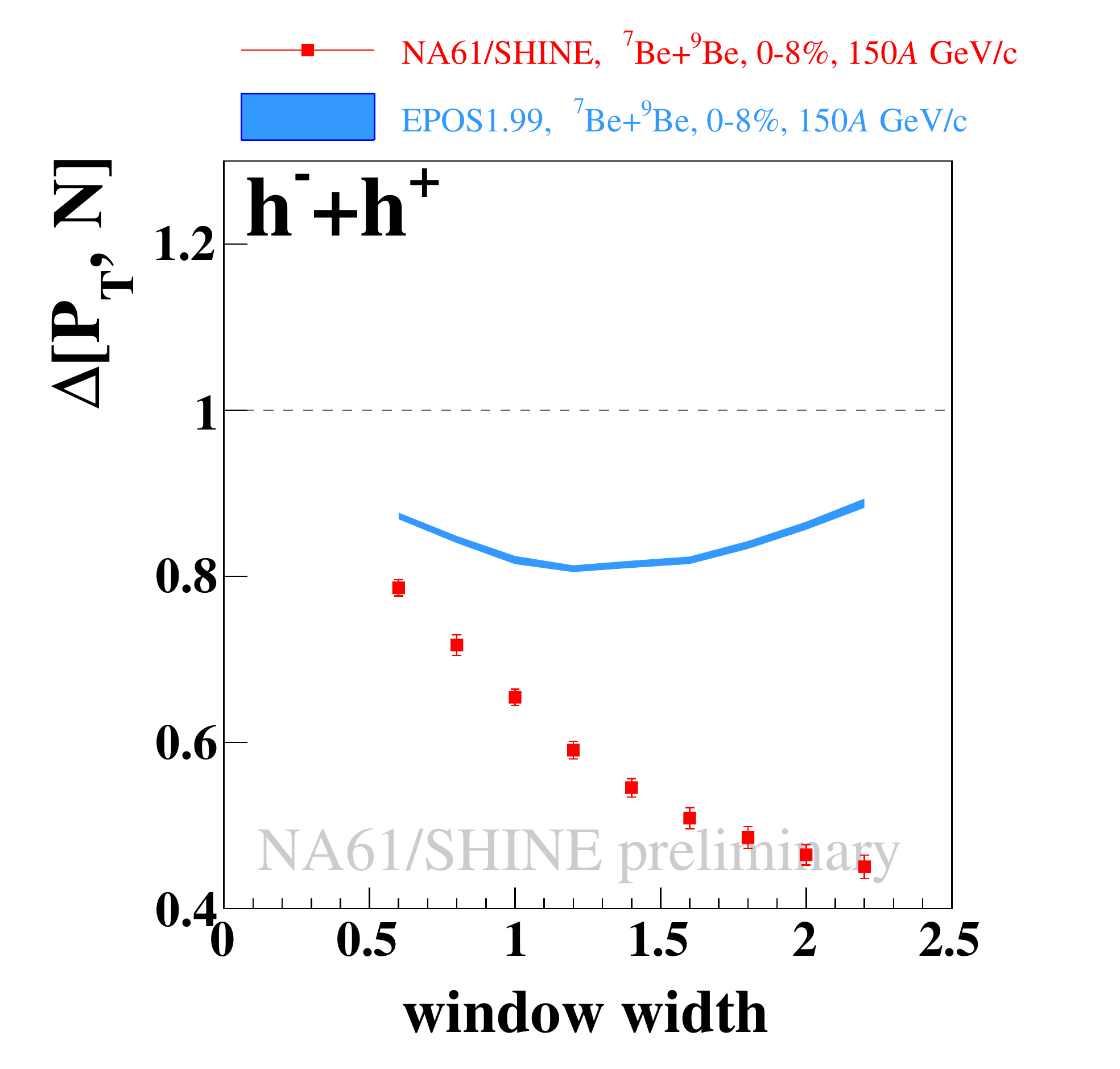}
\end{minipage}
\begin{minipage}{12pc}
\includegraphics[width=12pc]{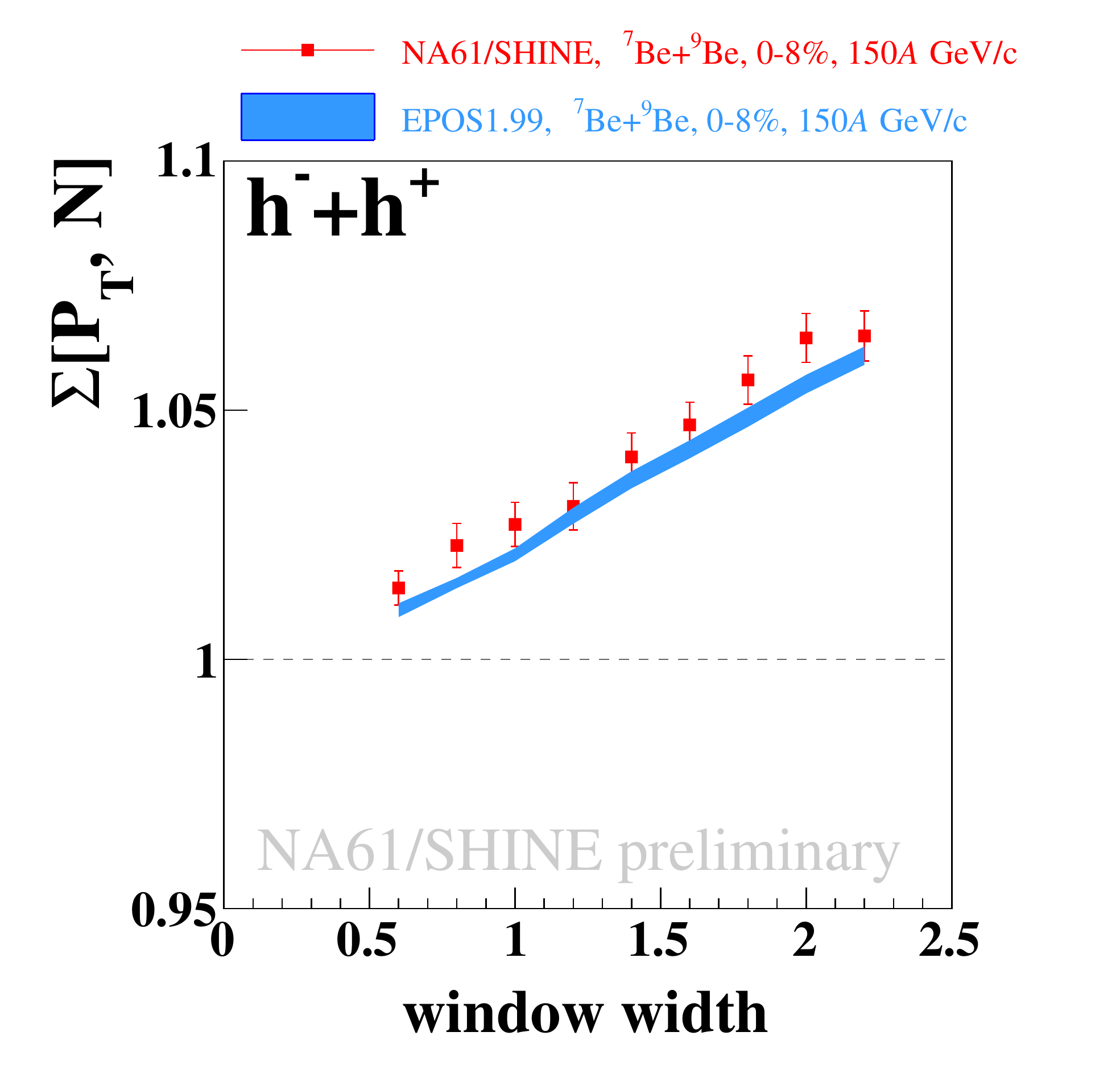}
\end{minipage}
\end{center}
\caption{\label{label1}Dependence of $\Delta[P_{T},N]$ (left) and $\Sigma[P_{T},N]$ (right) on the width of the pseudorapidity window. Red squares are preliminary NA61/SHINE measurements for 0-8$\%$ Be+Be collisions at 150{\it A} GeV/c. Blue band represents the EPOS1.99 model predictions.}
\end{figure}

Figure \ref{label1} shows preliminary results for the dependence of $\Delta[P_{T},N]$ and $\Sigma[P_{T},N]$ on the width of the pseudorapidity interval and comparisons to the EPOS1.99 model \cite{Pierog:2009zt} predictions. Both quantities change monotonously for the data in contrast to the EPOS1.99 results for $\Delta[P_{T},N]$ which show a minimum for intermediate width and lie significantly above the measurements. Another observation is that for smaller windows these quantities approach unity (independent particle production limit) as the number of particles in the interval gets small. Moreover, the inequalities $\Delta[P_{T},N]<1$ and $\Sigma[P_{T},N]\geq 1$, previously observed for all studied systems at all collision energies \cite{Andronov:2016ddd}, also hold when modifying the width of the pseudorapidity window of the measurement.  In general, no traces of the possible critical point of strongly interacting matter are visible.  

\section{Proton density fluctuations: intermittency}
In the grand canonical ensemble the correlation length diverges at the critical point and the system becomes scale invariant~\cite{Satz:1989vj}. In a pure critical system, intermittency in transverse momentum space can be revealed by the scaling of the Second Scaled Factorial Moments of protons as a function of bin size~\cite{BIALAS1986703}. For that purpose, a region of transverse momentum space is partitioned into $M\times M$ equal-size bins and these moments are defined as:
\beq\label{ssfm}
F_{2}(M)=\frac{\langle\frac{1}{M^{2}}\sum_{i=1}^{M^2}n_{i}(n_{i}-1) \rangle}{{\langle \frac{1}{M^{2}}\sum_{i=1}^{M^2}n_{i}\rangle}^{2}}
\eeq
where $n_{i}$ is the number of particles in the i-th bin. If the system exhibits critical fluctuations, $F_{2}(M)$ is expected to scale with M, for large values of M, as a power-law $F_{2}(M)\sim M^{2\phi_{2}}$, where $\phi_{2}$ is the intermittency index.

Intermittency analysis was performed for NA61/SHINE data on Be+Be collisions (0-10$\%$) at 150{\it A} GeV/{\it c} beam momentum. No significant intermittency effect was observed, with an upper limit for protons obeying critical fluctuations being of the order of 0.3$\%$. This limit was estimated based on the Critical Monte Carlo model predictions~\cite{Antoniou:2000ms}. For more details see Ref.~\cite{Davis_QM}.


\section{Acknowledgments}
This work is supported by the Russian Science Foundation under grant 17-72-20045.





\bibliographystyle{elsarticle-num}
\bibliography{eaReferences}







\end{document}